%% file: main.tex
\documentclass[a4paper]{article}

\usepackage{INTERSPEECH2020}
\usepackage{multirow}
\usepackage{breqn}
\usepackage{here}
\usepackage{url}

\newcommand{\argmax}{\mathop{\rm arg~max}\limits}

\title{Mask CTC: Non-Autoregressive End-to-End ASR with CTC and Mask Predict}
\name{Yosuke Higuchi$^{1, 2}$, Shinji Watanabe$^1$, Nanxin Chen$^1$, Tetsuji Ogawa$^2$, Tetsunori Kobayashi$^2$}

\address{
    $^1$ Center for Language and Speech Processing, Johns Hopkins University, Baltimore, USA \\
    $^2$ Department of Communications and Computer Engineering, Waseda University, Tokyo, Japan
  }
\email{higuchi@pcl.cs.waseda.ac.jp}

\begin{document}
\maketitle

\input{subtex/00_abstract}

\input{subtex/01_introduction}

\input{subtex/02_nar_asr}

\input{subtex/03_experiments}

\input{subtex/04_conclusions}

\input{subtex/bib}

\end{document}

%% file: subtex/00_abstract.tex
\begin{abstract}
We present Mask CTC, 
a novel \textit{non-autoregressive} end-to-end automatic speech recognition (ASR) framework, which generates a sequence by refining outputs of the connectionist temporal classification (CTC).
Neural sequence-to-sequence models are usually \textit{autoregressive}: 
each output token is generated by conditioning on previously generated tokens, 
at the cost of requiring as many iterations as the output length.
On the other hand, 
non-autoregressive models can simultaneously generate tokens within a constant number of iterations, 
which results in significant inference time reduction and 
better suits end-to-end ASR model for real-world scenarios.
In this work, 
Mask CTC model is trained using a Transformer encoder-decoder with joint training of mask prediction and CTC.
During inference, 
the target sequence is initialized with the greedy CTC outputs and 
low-confidence tokens are masked based on the CTC probabilities.
Based on the conditional dependence between output tokens, 
these masked low-confidence tokens are then predicted conditioning on the high-confidence tokens.
Experimental results on different speech recognition tasks show that 
Mask CTC outperforms the standard CTC model (e.g., 17.9\% $\rightarrow$ 12.1\% WER on WSJ) and 
approaches the autoregressive model, requiring much less inference time using CPUs (0.07 RTF in Python implementation).
All of our codes are publicly available at \url{https://github.com/espnet/espnet}.

\end{abstract}
\noindent\textbf{Index Terms}: non-autoregressive, connectionist temporal classification, transformer, end-to-end speech recognition

%% file: subtex/01_introduction.tex
\section{Introduction}
Owing to the rapid development of neural sequence-to-sequence modeling~\cite{sutskever2014sequence, bahdanau2014neural}, 
deep neural network (DNN)-based end-to-end automatic speech recognition (ASR) systems have become almost as effective as the traditional hidden Markov model-based systems~\cite{chiu2018state, luscher2019rwth, karita2019a}.
Various models and approaches have been proposed for improving the performance of the \textit{autoregressive} (AR) end-to-end ASR model with the encoder-decoder architecture based on recurrent neural networks (RNNs)~\cite{chorowski2015attention, chan2016listen, kim2017joint} and Transformers~\cite{vaswani2017attention, dong2018speech, karita2019improving}.

Contrary to the autoregressive framework, 
\textit{non-autoregressive} (NAR) sequence generation has attracted  attention, including the revisitation of connectionist temporal classification (CTC)~\cite{graves2006connectionist, libovicky2018end} and the growing interest for non-autoregressive Transformer (NAT)~\cite{gu2017non}.
While the autoregressive model requires $L$ iterations to generate an $L$-length target sequence, a non-autoregressive model costs a constant number of iterations $K (\ll L)$, independent on the length of the target sequence.
Despite the limitation in this decoding iteration, some recent studies in neural machine translation have successfully shown the effectiveness of the non-autoregressive models, 
performing comparable results to the autoregressive models.
Different types of non-autoregressive models have been proposed based on 
the iterative refinement decoding~\cite{lee2018deterministic}, 
insert or edit-based sequence generation~\cite{stern2019insertion, gu2019levenshtein}, 
masked language model objective~\cite{ghazvininejad2019mask, ghazvininejad2020semi, saharia2020non}, and
generative flow~\cite{ma2019flowseq}.

Some attempts have also been made to realize the non-autoregressive model in speech recognition.
CTC introduces a frame-wise latent alignment to represent the alignment between the input speech frames and the output tokens~\cite{graves2014towards}.
While CTC makes use of dynamic programming to efficiently calculate the most probable alignment, 
the strong conditional independence assumption between output tokens results in poor performance compared to the autoregressive models~\cite{battenberg2017exploring}.
On the other hand, \cite{chen2019non} trains a Transformer encoder-decoder in a mask-predict manner~\cite{ghazvininejad2019mask}: 
target tokens are randomly masked and predicted conditioning on the unmasked tokens and the input speech.
To generate the output sequence in parallel during inference, 
the target sequence is initialized as all masked tokens and the output length is predicted by finding the position of the end-of-sequence token.
However with this prediction of the output length, 
the model is known to be vulnerable to the output sequence with a long length. 
At the beginning of the decoding, 
the model is likely to make more mistakes in predicting long masked sequence,
propagating the error to the later decoding steps.
\cite{chan2020imputer} proposes Imputer, which performs the mask prediction in CTC's latent alignments to get rid of the output length prediction.
However, unlike the mask-predict, Imputer requires more calculations in each interaction, which is proportional to the square of the input length $T (\gg L)$ in the self-attention layer, and the total computational cost can be very large.

Our work aims to obtain a non-autoregressive end-to-end ASR model, 
which generates the sequence in token-level with low computational cost.
The proposed Mask CTC framework trains a Transformer encoder-decoder model with both CTC and mask-predict objectives. 
During inference, 
the target sequence is initialized with the greedy CTC outputs and low-confidence tokens are masked based on the CTC probabilities.
The masked low-confidence tokens are predicted conditioning on the high-confidence tokens not only in the past but also in the future context.
The advantages of Mask CTC are summarized as follows.

\noindent\textbf{No requirement for output length prediction:} 
Predicting the output token length from input speech is rather challenging 
because the length of the input utterances varies greatly depending on the speaking rate or the duration of silence.
By initializing the target sequence with the CTC outputs, 
Mask CTC does not have to care about predicting the output length at the beginning of the decoding.

\noindent\textbf{Accurate and fast decoding:} 
We observed that the results of CTC outputs themselves are quite accurate.
Mask CTC does not only retain the correct tokens in the CTC outputs but also recovers the output errors by considering the entire context.
Token-level iterative decoding with a small number of masks makes the model well-suited for 
the usage in real scenarios.

%% file: subtex/02_nar_asr.tex
\section{Mask CTC framework}
The objective of end-to-end ASR is to model the joint probability of a $L$-length output sequence $Y = \{ y_l \in \mathcal{V} | l=1,...,L \}$ given a $T$-length input sequence $X = \{\bm{\mathrm{x}}_t \in \mathbb{R}^D| t=1,...,T\}$. Here, $y_l$ is an output token at position $l$ in the vocabulary $\mathcal{V}$ and $\bm{\mathrm{x}}_t$ is a $D$-dimensional acoustic feature at frame $t$.

The following subsections first explain a conventional autoregressive framework based on attention-based encoder-decoder and CTC.
Then a non-autoregressive model trained with mask prediction is explained and finally, the  proposed Mask CTC decoding method is introduced.

\subsection{Attention-based encoder-decoder}
Attention-based encoder-decoder models the joint probability of $Y$ given $X$ by factorizing the probability based on the probabilistic left-to-right chain rule as follows:
\begin{equation}
    \label{eq:p_att}
     P_{\mathrm{att}} (Y | X) = \prod_{l=1}^{L} P_{\mathrm{att}} (y_l | y_{<l}, X).
\end{equation}
The model estimates the output token $y_l$ at each time-step conditioning on previously generated tokens $y_{<l}$ in an autoregressive manner. 
In general, during training, 
the ground truth tokens are used for the history tokens $y_{<l}$ and 
during inference, the predicted tokens are used.

\subsection{Connectionist temporal classification}
CTC predicts a frame-level alignment between the input sequence $X$ and the output sequence $Y$ by introducing a special \texttt{<blank>} token.
The alignment $A=\{ a_t \in \mathcal{V} \cup \{\texttt{<blank>}\} | t=1,...,T\}$ is predicted with the conditional independence assumption between the output tokens as follows:
\begin{equation}
    P_{\mathrm{ctc}} (A | X) = \prod_{t=1}^{T} P_{\mathrm{ctc}} (a_t | X).
\end{equation}
Considering the probability distribution over all possible alignments, 
CTC models the joint probability of $Y$ given $X$ as follows:
\begin{equation}
    \label{eq:p_ctc}
    P_{\mathrm{ctc}} (Y | X) = \sum_{A \in \beta^{-1} (Y)} P_{\mathrm{ctc}} (A | X),
\end{equation}
where $\beta^{-1} (Y)$ returns all possible alignments compatible with $Y$.
The summation of the probabilities for all of the alignments can be computed efficiently by using dynamic programming.

To achieve robust alignment training and fast convergence, 
an end-to-end ASR model based on an attention-based encoder-decoder framework is trained with CTC~\cite{kim2017joint, karita2019improving}. 
The objective of the autoregressive joint CTC-attention model is defined as follows by combining Eq. (\ref{eq:p_att}) and Eq. (\ref{eq:p_ctc}):
\begin{equation}
    \label{eq:ctc-att}
    \mathcal{L}_{\mathrm{AR}} = \lambda \log P_{\mathrm{ctc}} (Y | X) + 
    (1 - \lambda) \log P_{\mathrm{att}} (Y | X), 
\end{equation}
where $\lambda$ is a tunable parameter.

\subsection{Joint CTC-CMLM non-autoregressive ASR}
Mask CTC adopts non-autoregressive speech recognition~\cite{chen2019non} based on a conditional masked language model (CMLM)~\cite{ghazvininejad2019mask}, where the model is trained to predict masked tokens in the target sequence~\cite{devlin2019bert}\footnote{Note that CMLM is used as an ASR decoder network conditioned on the encoder output as well, and it is different from an external language model often used in shallow fusion during decoding.}.
Taking advantages of Transformer's parallel computation~\cite{vaswani2017attention}, 
CMLM can predict any arbitrary subset of masked tokens in the target sequence by attending to the entire sequence including tokens in the past and the future.

CMLM predicts a set of masked tokens $Y_{\mathrm{mask}}$ conditioning on the input sequence $X$ and observed (unmasked) tokens $Y_{\mathrm{obs}}$ as follows:
\begin{equation}
      P_{\mathrm{cmlm}} (Y_{\mathrm{mask}} | Y_{\mathrm{obs}}, X) = 
      \prod_{y \in Y_{\mathrm{mask}}} P_{\mathrm{cmlm}} (y | Y_{\mathrm{obs}}, X),
      \label{eq:p_cmlm}
\end{equation}
where $Y_{\mathrm{obs}}  = Y \setminus Y_{\mathrm{mask}}$.
During training, 
the ground truth tokens are randomly replaced by a special \texttt{<MASK>} token and CMLM is trained to predict the original tokens conditioning on the input sequence $X$ and the unmasked tokens $Y_{\mathrm{obs}}$.
The number of tokens to be masked is sampled from a uniform distribution between 1 to $L$ as in~\cite{ghazvininejad2019mask}.
During inference, 
the target sequence is gradually generated in a constant number of iterations $K$ by the iterative decoding algorithm~\cite{ghazvininejad2019mask}, 
which repeatedly masks and predicts the subset of the target sequence.

We observed that applying the original CMLM to non-autoregressive speech recognition shows poor performance, having the problem of skipping and repeating the output tokens.
To deal with this, 
we found that jointly training with CTC similar to \cite{kim2017joint} provides the model with absolute positional information (conditional independence) explicitly and improves the model performance reasonably well.
With the CTC objective from Eq. (\ref{eq:p_ctc}) and Eq. (\ref{eq:p_cmlm}), the objective of joint CTC-CMLM training for non-autoregressive ASR model is defined as follows:
\begin{dmath}
    \label{eq:ctc-cmlm}
    \mathcal{L}_{\mathrm{NAR}} = 
    \gamma \log P_{\mathrm{ctc}} (Y | X) + 
    (1 - \gamma) \log P_{\mathrm{cmlm}} (Y_{\mathrm{mask}} | Y_{\mathrm{obs}}, X),
\end{dmath}
where $\gamma$ is a tunable parameter.

\subsection{Mask CTC decoding} \label{ssec:CTC-mask}
\begin{figure}[t]
    \centering
    \hspace{-35pt}
    \includegraphics[width=0.7 \columnwidth]{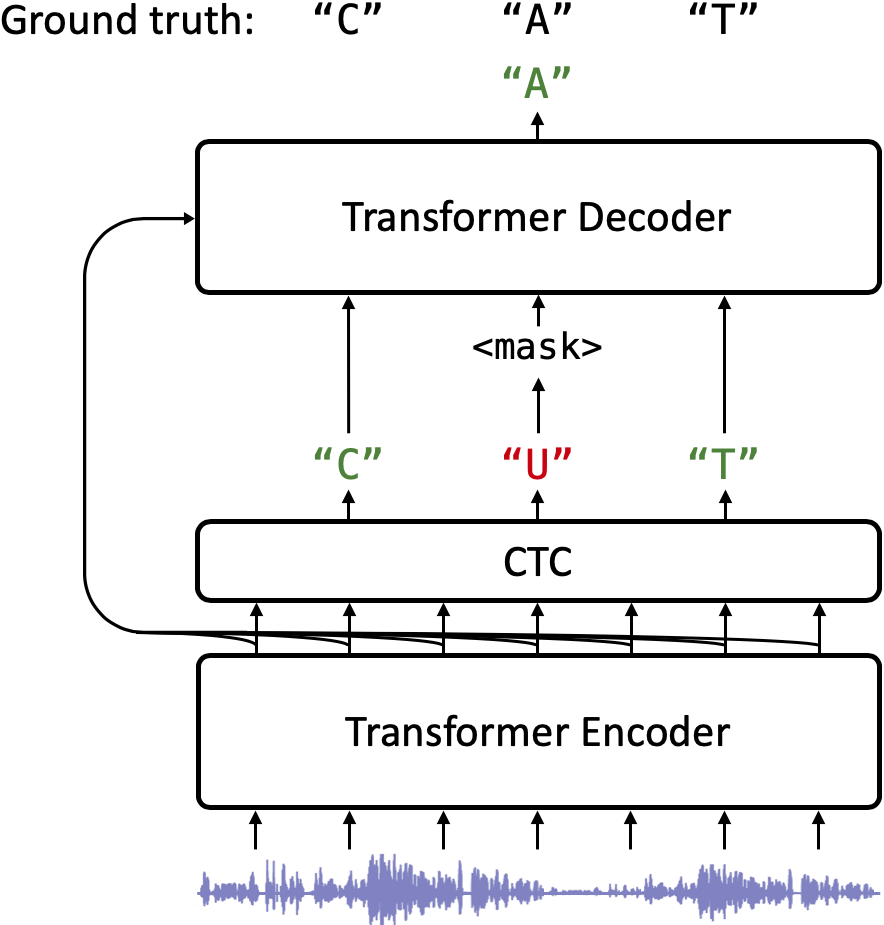}
    \vspace{-4pt}
    \caption{Overview of Mask CTC predicting ``CAT'' based on CTC outputs. The model is trained with the joint CTC and mask-predict objectives. During inference, the target sequence is initialized with the greedy CTC outputs and low-confidence tokens are masked based on the CTC probabilities. The masked low-confidence tokens are predicted conditioning on the high-confidence tokens.}
    \vspace{-12pt}
    \label{fig:proposed_decoding}
\end{figure}

Non-autoregressive models must know the length of the output sequence
to predict the entire sequence in parallel.
For example, in the beginning of the CMLM decoding, 
the output length must be given to initialize the target sequence with the masked tokens.
To deal with this problem, 
in machine translation, 
the output length is predicted by training a fertility model~\cite{gu2017non} or 
introducing a special \texttt{<LENGTH>} token in the encoder~\cite{ghazvininejad2019mask}.
In speech recognition, however, 
due to the different characteristics between the input acoustic signals and the output linguistic symbols, 
it appeared that predicting the output length is rather challenging, 
e.g., the length of the input utterances of the same transcription varies greatly depending on the speaking rate or the duration of silence.
\cite{chen2019non} simply makes the decoder to predict the position of \texttt{<EOS>} token to deal with the output length.
However, they analyzed that this prediction is vulnerable to the output sequence with a long length because the model is likely to make more mistakes in predicting a long masked sequence and the error is propagated to the later decoding, which degrades the recognition performance.
To compensate this problem, 
they use beam search with CTC and a language model to obtain the reasonable performance, 
which leads to a slow down of the overall decoding speed, 
making the advantage of non-autoregressive framework less effective.

To tackle this problem regarding the initialization of the target sequence, 
we consider using the CTC outputs as the initial sequence for decoding.
Figure~\ref{fig:proposed_decoding} shows the decoding of CTC Mask based on the inference of CTC. 
CTC outputs are first obtained through a single calculation of the encoder and the decoder works as to refine the CTC outputs by attending to the whole sequence.

In this work, we use ``greedy'' result of CTC $\hat{Y} = \{ \hat{y}_l \in \beta(A) | l=1,...,L' \}$, which is obtained without using prefix search~\cite{graves2006connectionist}, to keep an inference algorithm non-autoregressive.
The errors caused by the conditional independence assumption are expected to be corrected using the CMLM decoder.
The posterior probability of $\hat{y}_l$ is approximately calculated by using the frame-level CTC probabilities as follows:
\begin{equation}
    \hat{P}(\hat{y}_l | X) = \max_{j} \Big( P(a_j \in A_l | X) \Big),
\end{equation}
where $A_l = \{a_j\}_j$ is the consecutive same alignments that corresponds to the aggregated token $\hat{y}_l$.
Then, a part of $\hat{Y}$ is masked-out based on a confidence using the probability $\hat{P}$ as follows:
\begin{eqnarray}
    \hat{Y}_{\mathrm{mask}} = \{ y_l \in \hat{Y} | \hat{P}(y_l | X) < P_{\mathrm{thres}} \},
    \label{eq:Y_hat_mask}
\end{eqnarray}
where $P_{\mathrm{thres}}$ is a threshold to decide whether the target token is masked or not.
Finally, $\hat{Y}_{\mathrm{mask}}$ is predicted conditioning on the high-confidence tokens $\hat{Y}_{\mathrm{obs}} = \hat{Y} \setminus \hat{Y}_{\mathrm{mask}}$ and the input sequence $X$ as in Eq. (5).

We also investigated applying one of the iterative decoding methods called easy-first~\cite{goldberg2010efficient, chen2019non}.
Starting with the masked CTC output,
the masked tokens are gradually predicted by a confidence based on the CMLM probability.
In the $n$-th decoding iteration, 
$\hat{y}_{l} \in \hat{Y}_{\mathrm{mask}}^{(n)}$ is updated as follows:
\begin{equation}
    \hat{y}_l= \argmax_{w} P_{\mathrm{cmlm}} (\hat{y}_l = w | Y_{\mathrm{obs}}^{(n)}, X).
\end{equation}
Top $C$ masked tokens with the highest probabilities are predicted in each iteration.
By defining $C = [L / K]$, 
the number of total decoding iterations can be controlled in a constant $K$ iterations.

With this proposed non-autoregressive training and decoding with Mask CTC, 
the model does not have to take care about predicting the output length.
Moreover, 
decoding by refining CTC outputs with the mask prediction is expected to compensate the errors come from the conditional independence assumption.

%% file: subtex/03_experiments.tex
\section{Experiments}
To evaluate the effectiveness Mask CTC, we conducted 
speech recognition experiments to compare different end-to-end ASR models using ESPnet~\cite{watanabe2018espnet}.
The performance of the models was evaluated based on character error rates (CERs) or word error rates (WERs) without relying on external language models.

\begin{table}[t]
    \centering
    \caption{Word error rates (WERs) and real time factor (RTF) for WSJ (English).}
    \vspace{-10pt}
    \label{tb:wsj}
    \scalebox{0.90}{
    \begin{tabular}{lcccc}
        \toprule
         \textbf{Model} & \textbf{Iterations} & \textbf{dev93} & \textbf{eval92} & \textbf{RTF} \\
        \midrule
        \textit{Autoregressive} & \\
        \quad CTC-attention~\cite{kim2017joint} & $L$ & 14.4 & 11.3 & 0.97 \\ 
        \qquad + beam search & $L$ & 13.5 & 10.9 & 4.62 \\ 
        \midrule
        \textit{Non-autoregressive} & \\
        \quad CTC & 1 & 22.2 & 17.9 & 0.03 \\ 
        \quad Mask CTC & \multirow{2}{*}{1}  & \multirow{2}{*}{16.3} & \multirow{2}{*}{12.9} & \multirow{2}{*}{0.03}  \\ 
        \qquad ($P_{\mathrm{thres}}=0.0$) \\
        \quad Mask CTC & 1 & 15.7 & 12.5 & 0.04 \\ 
        \quad Mask CTC & 5 & 15.5 & 12.2 & 0.05 \\ 
        \quad Mask CTC & 10 & 15.5 & \bf{12.1} & 0.07 \\ 
        \quad Mask CTC & \#mask & \bf{15.4} & \bf{12.1} & 0.13 \\ 
        \midrule
        \quad CTC~\cite{chan2020imputer} & 1 & -- & 15.2 & -- \\
        \quad Imputer (IM)~\cite{chan2020imputer} & 8 & -- & 16.5 & -- \\
        \quad Imputer (DP)~\cite{chan2020imputer} & 8 & -- & 12.7 & -- \\
        \bottomrule
    \end{tabular}}
    \vspace{-7pt}
\end{table}

\begin{table}[t]
    \centering
    \caption{Word error rates (WERs) for Voxforge (Italian).}
    \vspace{-10pt}
    \label{tb:voxforge}
    \scalebox{0.90}{
    \begin{tabular}{lcccc}
        \toprule
        \textbf{Model} & \textbf{Iterations} & \textbf{Dev} & \textbf{Test} \\
        \midrule
        \textit{Autoregressive} & \\
        \quad CTC-attention~\cite{kim2017joint} & $L$ & 35.5 & 35.5 \\ 
        \qquad + beam search & $L$ & 35.4 & 35.7 \\ 
        \midrule
        \textit{Non-autoregressive} & \\
        \quad CTC & 1 & 53.8 & 56.1 \\
        \quad Mask CTC ($P_{\mathrm{thres}}=0.0$) & 1 & 41.6 & 40.2 \\
        \quad Mask CTC & 1 & 41.3 & 40.1 \\
        \quad Mask CTC & 5 & 40.7 & 39.4 \\
        \quad Mask CTC & 10 & 40.5 & 39.2 \\
        \quad Mask CTC & \#mask & \bf{40.4} & \bf{39.0} \\
        \bottomrule
    \end{tabular}}
    \vspace{-14pt}
\end{table}

\begin{figure*}[t]
    \centering
    \includegraphics[width=1.0\linewidth]{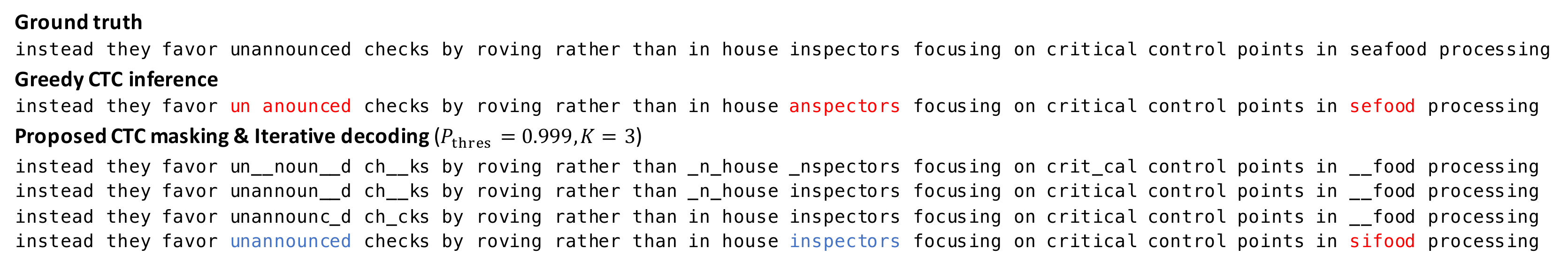}
    \vspace{-22pt}
    \caption{Decoding example for utterance 443c040i in WSJ eval92. The target sequence is initialized as the CTC outputs and some tokens are replaced with masks (``\_'') based on the CTC confidence. Then, the masked tokens are iteratively predicted conditioning on the other unmasked tokens. Red indicates characters with errors and blue indicates ones recovered by Mask CTC decoding.}
    \label{fig:decoding}
    \vspace{-15pt}
\end{figure*}

\subsection{Datasets}
The experiments were carried out using three tasks with different languages and amounts of training data: the 81 hours Wall Street Journal (WSJ) in English~\cite{paul1992design}, the 581 hours Corpus of Spontaneous Japanese (CSJ) in Japanese~\cite{maekawa2003corpus} and the 16 hours Voxforge in Italian~\cite{voxforge}.
For the network inputs, we used 80 mel-scale filterbank coefficients with three-dimensional pitch features and applied SpecAugment~\cite{park2019specaugment} during model training.
For the tokenization of the target, we used characters: 
Latin alphabets for English and Italian, and
Japanese syllable characters (Kana) and Chinese characters (Kanji) for Japanese.

\subsection{Experimental setup}
For experiments in all of the tasks, 
we adopted the same encoder-decoder architecture as~\cite{karita2019improving}, 
which consists of Transformer self-attention layers with 4 attention heads, 256 hidden units, and 2048 feed-forward inner dimension size.
The encoder included 12 self-attention layers with convolutional layers for downsampling and the decoder was 6 self-attention layers.
With the mask-predict objective, 
the convergence 
for training the Mask CTC model required more epochs (about 200 -- 500) than the autoregressive models (about 50 -- 100).
The final autoregressive model was obtained by averaging the model parameters of the last 10 epochs as in~\cite{karita2019a}.
For Mask CTC model, 
we found that the model performance was significantly improved by 
averaging the model parameters of 10 -- 30 epochs with the top validation accuracies.
For the threshold $P_{\mathrm{thres}}$ in Eq. (\ref{eq:Y_hat_mask}), 
we used 0.999, 0.999, and 0.9 for WSJ, Voxforge, and CSJ, respectively.
For all of the tasks, 
the loss weights $\lambda$ and $\gamma$ in Eq. (\ref{eq:ctc-att}) and Eq. (\ref{eq:ctc-cmlm}) were set to 0.3 and 0.3, respectively.

\subsection{Evaluated models}
\begin{itemize}
    \setlength{\parskip}{0cm}
    \setlength{\itemsep}{0.0cm}
    \item {\bf CTC-attention}: An autoregressive model trained with the joint CTC-attention objective as in Eq. (\ref{eq:ctc-att}). During inference, the joint CTC-attention decoding is applied with beam search~\cite{hori2017joint}.
    \item {\bf CTC}: A non-autoregressive model simply trained with the CTC objective.
    \item {\bf Mask CTC}: A non-autoregressive model trained with the joint CTC-CMLM objective as in Eq. (\ref{eq:ctc-cmlm}). During inference, the proposed decoding based on masking the CTC outputs (explianed in Section \ref{ssec:CTC-mask}) is applied. Note that when $P_{\mathrm{thres}} = 0.0$ in Eq. (\ref{eq:Y_hat_mask}), the greedy output of CTC was used as a decoded result.
    \vspace{-5pt}
\end{itemize}

\subsection{Results}
\begin{table}[t]
    \centering
    \caption{Character error rates (CERs) and sentence error rates (SERs) for CSJ (Japanese).}
    \vspace{-10pt}
    \label{tb:csj}
    \scalebox{0.82}{
    \begin{tabular}{lcccccc}
        \toprule
        \multirow{2}{*}{\textbf{Model}} & \multicolumn{2}{c}{\textbf{Eval1}} & \multicolumn{2}{c}{\textbf{Eval2}} & \multicolumn{2}{c}{\textbf{Eval3}} \\
        & \textbf{CER} & \textbf{SER} & \textbf{CER} & \textbf{SER} & \textbf{CER} & \textbf{SER} \\
        \midrule
        \textit{Autoregressive} & \\
        \quad CTC-attention~\cite{kim2017joint} & 6.37 & 57.0 & 4.76 & 53.7 & 5.40 & 39.6 \\ 
        \qquad + beam search & 6.21 & 56.8 & 4.50 & 53.4 & 5.15 & 40.1 \\ 
        \midrule
        \textit{Non-autoregressive} & \\
        \quad CTC & \bf{6.51} & 59.7 & 4.71 & 59.5 & 5.49 & 44.5 \\
        \quad Mask CTC & \multirow{2}{*}{6.56} & \multirow{2}{*}{60.3} & \multirow{2}{*}{4.69} & \multirow{2}{*}{57.0} & \multirow{2}{*}{4.97} & \multirow{2}{*}{41.9} \\
        \qquad ($P_{\mathrm{thres}}=0.0$) \\
        \quad Mask CTC ($K=5$) & 6.56 & \bf{58.7} & \bf{4.57} & \bf{55.5} & \bf{4.96} & \bf{40.7} \\
        \bottomrule
    \end{tabular}}
    \vspace{-12pt}
\end{table}

Table~\ref{tb:wsj} shows the results for WSJ based on WERs and real time factors (RTFs)
that were measured for decoding eval92 with Intel(R) Core(TM), i9-7980XE, 2.60GHz.
By comparing the results for non-autoregressive models, 
we can see that the greedy CTC outputs of Mask CTC outperformed the simple CTC model by training with the mask-predict objective.
By applying the refinement based on the proposed CTC masking, 
the model performance was steadily improved.
The performance was further improved by increasing the number of decoding iterations and 
it resulted in the best performance with \#mask iterations, which means one mask is predicted in each iteration.
The results of Mask CTC are reasonable comparing to the results of prior work~\cite{chan2020imputer}.
Our models also approached the results of autoregressive models from the initial CTC result.
In terms of the decoding speed measured in RTF, 
Mask CTC is, at most, 116 times faster than the autoregressive models.
Since most of the CTC outputs are fairly accurate and the number of masks are quite small, 
there was not so much degradation in the speed as the number of the decoding iterations was increased.

Figure~\ref{fig:decoding} shows an example decoding process of a sample in the WSJ evaluation set. 
Here, we can see that the CTC outputs include errors mainly coming from substitution errors due to the incomplete word spelling.
By applying Mask CTC decoding, 
the spelling errors were successfully recovered by considering the conditional dependence between characters in word-level.
However, 
as can be seen in the error for ``sifood,'' 
Mask CTC cannot recover errors derived from character-level insertion or deletion errors 
because the length allocated to each word is fixed by the CTC outputs.

Table~\ref{tb:voxforge} shows WERs for Voxforge.
Mask CTC yielded better scores than the standard CTC model as the similar results to WSJ, 
demonstrating that our model can be adopted to other languages with a relatively small amount of training data.

Table~\ref{tb:csj} shows character error rates (CERs) and sentence error rates (SERs) for CSJ.
While Mask CTC performed quite close or even better CERs than the autoregressive model, 
the results showed a little improvement from the simple CTC model, 
compared to the results of the aforementioned tasks.
Since Japanese includes a large number of characters and the characters themselves often form a certain word,
the simple CTC model seemed to be dealing with the short dependence between the characters reasonably well, 
performing almost the same scores without applying Mask CTC.
However, 
when we look at the results in sentence-level, 
we observed some clear improvements for all of the evaluation sets, 
again showing that our model effectively recovers the CTC errors by considering the conditional dependence.

These experimental results on different tasks indicate that 
Mask CTC framework is especially effective on languages having tokens with a small unit (i.e., Latin alphabet and other phonemic scripts).
It is our future work for investigating the effectiveness when we use byte pair encodings (BPEs)~\cite{sennrich2016neural} for the languages with such a small unit.


%% file: subtex/04_conclusions.tex
\section{Conclusions}
This paper proposed Mask CTC, 
a novel non-autoregressive end-to-end speech recognition framework, which generates a sequence by refining the CTC outputs based on mask prediction.
During inference, the target sequence was initialized with the greedy CTC outputs and low-confidence masked tokens were iteratively refined conditioning on the other unmasked tokens and input speech features.
The experimental comparisons demonstrated that Mask CTC outperformed the standard CTC model while maintaining the decoding speed fast.
Mask CTC approached the results of autoregressive models; 
especially for CSJ, they were comparable or even better.
Our future plan is to reduce the gap of masking strategies between training using random masking and inference using CTC outputs.
Furthermore, 
we plan to explore the integration of external language models (e.g., BERT~\cite{devlin2019bert}) in Mask CTC framework.

%% file: subtex/bib.tex
\bibliographystyle{IEEEtran}
\bibliography{refs}

%% file: main.bbl
\begin{thebibliography}{10}
\providecommand{\url}[1]{#1}
\csname url@samestyle\endcsname
\providecommand{\newblock}{\relax}
\providecommand{\bibinfo}[2]{#2}
\providecommand{\BIBentrySTDinterwordspacing}{\spaceskip=0pt\relax}
\providecommand{\BIBentryALTinterwordstretchfactor}{4}
\providecommand{\BIBentryALTinterwordspacing}{\spaceskip=\fontdimen2\font plus
\BIBentryALTinterwordstretchfactor\fontdimen3\font minus
  \fontdimen4\font\relax}
\providecommand{\BIBforeignlanguage}[2]{{%
\expandafter\ifx\csname l@#1\endcsname\relax
\typeout{** WARNING: IEEEtran.bst: No hyphenation pattern has been}%
\typeout{** loaded for the language `#1'. Using the pattern for}%
\typeout{** the default language instead.}%
\else
\language=\csname l@#1\endcsname
\fi
#2}}
\providecommand{\BIBdecl}{\relax}
\BIBdecl

\bibitem{sutskever2014sequence}
I.~Sutskever, O.~Vinyals, and Q.~V. Le, ``Sequence to sequence learning with
  neural networks,'' in \emph{Proceedings of Advances in Neural Information
  Processing Systems (NeurIPS)}, 2014.

\bibitem{bahdanau2014neural}
D.~Bahdanau, K.~Cho, and Y.~Bengio, ``Neural machine translation by jointly
  learning to align and translate,'' in \emph{Proceedings of International
  Conference on Learning Representations (ICLR)}, 2015.

\bibitem{chiu2018state}
C.-C. Chiu, T.~N. Sainath, Y.~Wu, R.~Prabhavalkar, P.~Nguyen, Z.~Chen,
  A.~Kannan, R.~J. Weiss, K.~Rao, E.~Gonina \emph{et~al.}, ``State-of-the-art
  speech recognition with sequence-to-sequence models,'' in \emph{Proceedings
  of IEEE International Conference on Acoustics, Speech and Signal Processing
  (ICASSP)}, 2018.

\bibitem{luscher2019rwth}
C.~L{\"u}scher, E.~Beck, K.~Irie, M.~Kitza, W.~Michel, A.~Zeyer,
  R.~Schl{\"u}ter, and H.~Ney, ``{RWTH} {ASR} systems for librispeech: Hybrid
  vs attention,'' in \emph{Proceedings of Annual Conference of the
  International Speech Communication Association (INTERSPEECH)}, 2019.

\bibitem{karita2019a}
S.~Karita, X.~Wang, S.~Watanabe, T.~Yoshimura, W.~Zhang, N.~Chen, T.~Hayashi,
  T.~Hori, H.~Inaguma, Z.~Jiang, M.~Someki, N.~Yalta, and R.~Yamamoto, ``A
  comparative study on {Transformer} vs {RNN} in speech applications,'' in
  \emph{Proceedings of IEEE Workshop on Automatic Speech Recognition and
  Understanding (ASRU)}, 2019.

\bibitem{chorowski2015attention}
J.~K. Chorowski, D.~Bahdanau, D.~Serdyuk, K.~Cho, and Y.~Bengio,
  ``Attention-based models for speech recognition,'' in \emph{Proceedings of
  Advances in Neural Information Processing Systems (NeurIPS)}, 2015.

\bibitem{chan2016listen}
W.~Chan, N.~Jaitly, Q.~Le, and O.~Vinyals, ``Listen, attend and spell: A neural
  network for large vocabulary conversational speech recognition,'' in
  \emph{Proceedings of IEEE International Conference on Acoustics, Speech and
  Signal Processing (ICASSP)}, 2016.

\bibitem{kim2017joint}
S.~Kim, T.~Hori, and S.~Watanabe, ``Joint {CTC}-attention based end-to-end
  speech recognition using multi-task learning,'' in \emph{Proceedings of IEEE
  International Conference on Acoustics, Speech and Signal Processing
  (ICASSP)}, 2017.

\bibitem{vaswani2017attention}
A.~Vaswani, N.~Shazeer, N.~Parmar, J.~Uszkoreit, L.~Jones, A.~N. Gomez,
  {\L}.~Kaiser, and I.~Polosukhin, ``Attention is all you need,'' in
  \emph{Proceedings of Advances in Neural Information Processing Systems
  (NeurIPS)}, 2017.

\bibitem{dong2018speech}
L.~Dong, S.~Xu, and B.~Xu, ``Speech-{Transformer}: a no-recurrence
  sequence-to-sequence model for speech recognition,'' in \emph{Proceedings of
  IEEE International Conference on Acoustics, Speech and Signal Processing
  (ICASSP)}, 2018.

\bibitem{karita2019improving}
S.~Karita, N.~E.~Y. Soplin, S.~Watanabe, M.~Delcroix, A.~Ogawa, and
  T.~Nakatani, ``Improving {Transformer}-based end-to-end speech recognition
  with connectionist temporal classification and language model integration,''
  in \emph{Proceedings of Annual Conference of the International Speech
  Communication Association (INTERSPEECH)}, 2019.

\bibitem{graves2006connectionist}
A.~Graves, S.~Fern{\'a}ndez, F.~Gomez, and J.~Schmidhuber, ``Connectionist
  temporal classification: labelling unsegmented sequence data with recurrent
  neural networks,'' in \emph{Proceedings of International Conference on
  Machine Learning (ICML)}, 2006.

\bibitem{libovicky2018end}
J.~Libovick{\'y} and J.~Helcl, ``End-to-end non-autoregressive neural machine
  translation with connectionist temporal classification,'' in
  \emph{Proceedings of Conference on Empirical Methods in Natural Language
  Processing (EMNLP)}, 2018.

\bibitem{gu2017non}
J.~Gu, J.~Bradbury, C.~Xiong, V.~O. Li, and R.~Socher, ``Non-autoregressive
  neural machine translation,'' \emph{Proceedings of International Conference
  on Learning Representations (ICLR)}, 2018.

\bibitem{lee2018deterministic}
J.~Lee, E.~Mansimov, and K.~Cho, ``Deterministic non-autoregressive neural
  sequence modeling by iterative refinement,'' in \emph{Proceedings of
  Conference on Empirical Methods in Natural Language Processing (EMNLP)},
  2018.

\bibitem{stern2019insertion}
M.~Stern, W.~Chan, J.~Kiros, and J.~Uszkoreit, ``Insertion {Transformer}:
  Flexible sequence generation via insertion operations,'' in \emph{Proceedings
  of International Conference on Machine Learning (ICML)}, 2019.

\bibitem{gu2019levenshtein}
J.~Gu, C.~Wang, and J.~Zhao, ``Levenshtein {Transformer},'' in
  \emph{Proceedings of Advances in Neural Information Processing Systems
  (NeurIPS)}, 2019.

\bibitem{ghazvininejad2019mask}
M.~Ghazvininejad, O.~Levy, Y.~Liu, and L.~Zettlemoyer, ``Mask-predict: Parallel
  decoding of conditional masked language models,'' in \emph{Proceedings of
  Conference on Empirical Methods in Natural Language Processing and
  International Joint Conference on Natural Language Processing
  (EMNLP-IJCNLP)}, 2019.

\bibitem{ghazvininejad2020semi}
M.~Ghazvininejad, O.~Levy, and L.~Zettlemoyer, ``Semi-autoregressive training
  improves mask-predict decoding,'' \emph{arXiv preprint arXiv:2001.08785},
  2020.

\bibitem{saharia2020non}
C.~Saharia, W.~Chan, S.~Saxena, and M.~Norouzi, ``Non-autoregressive machine
  translation with latent alignments,'' \emph{arXiv preprint arXiv:2004.07437},
  2020.

\bibitem{ma2019flowseq}
X.~Ma, C.~Zhou, X.~Li, G.~Neubig, and E.~Hovy, ``{FlowSeq}: Non-autoregressive
  conditional sequence generation with generative flow,'' in \emph{Proceedings
  of Conference on Empirical Methods in Natural Language Processing and
  International Joint Conference on Natural Language Processing
  (EMNLP-IJCNLP)}, 2019.

\bibitem{graves2014towards}
A.~Graves and N.~Jaitly, ``Towards end-to-end speech recognition with recurrent
  neural networks,'' in \emph{Proceedings of International Conference on
  Machine Learning (ICML)}, 2014.

\bibitem{battenberg2017exploring}
E.~Battenberg, J.~Chen, R.~Child, A.~Coates, Y.~G.~Y. Li, H.~Liu, S.~Satheesh,
  A.~Sriram, and Z.~Zhu, ``Exploring neural transducers for end-to-end speech
  recognition,'' in \emph{Proceedings of IEEE Automatic Speech Recognition and
  Understanding Workshop (ASRU)}, 2017.

\bibitem{chen2019non}
N.~Chen, S.~Watanabe, J.~Villalba, and N.~Dehak, ``Non-autoregressive
  {Transformer} automatic speech recognition,'' \emph{arXiv preprint
  arXiv:1911.04908}, 2019.

\bibitem{chan2020imputer}
W.~Chan, C.~Saharia, G.~Hinton, M.~Norouzi, and N.~Jaitly, ``Imputer: Sequence
  modelling via imputation and dynamic programming,'' \emph{arXiv preprint
  arXiv:2002.08926}, 2020.

\bibitem{devlin2019bert}
J.~Devlin, M.-W. Chang, K.~Lee, and K.~Toutanova, ``{BERT}: Pre-training of
  deep bidirectional transformers for language understanding,'' in
  \emph{Proceedings of Conference of the North American Chapter of the
  Association for Computational Linguistics: Human Language Technologies
  (NAACL-HLT)}, 2019.

\bibitem{goldberg2010efficient}
Y.~Goldberg and M.~Elhadad, ``An efficient algorithm for easy-first
  non-directional dependency parsing,'' in \emph{Proceedings of Conference of
  the North American Chapter of the Association for Computational Linguistics:
  Human Language Technologies (NAACL-HLT)}, 2010.

\bibitem{watanabe2018espnet}
S.~Watanabe, T.~Hori, S.~Karita, T.~Hayashi, J.~Nishitoba, Y.~Unno, N.~{Enrique
  Yalta Soplin}, J.~Heymann, M.~Wiesner, N.~Chen, A.~Renduchintala, and
  T.~Ochiai, ``{ESPnet}: End-to-end speech processing toolkit,'' in
  \emph{Proceedings of Annual Conference on the International Speech
  Communication Association (INTERSPEECH)}, 2018.

\bibitem{paul1992design}
D.~B. Paul and J.~M. Baker, ``The design for the wall street journal-based
  {CSR} corpus,'' in \emph{Proceedings of Workshop on Speech and Natural
  Language}, 1992.

\bibitem{maekawa2003corpus}
K.~Maekawa, ``Corpus of spontaneous {Japanese}: Its design and evaluation,'' in
  \emph{Proceedings of ISCA \& IEEE Workshop on Spontaneous Speech Processing
  and Recognition}, 2003.

\bibitem{voxforge}
``Voxforge,'' \url{http://www.voxforge.org}.

\bibitem{park2019specaugment}
D.~S. Park, W.~Chan, Y.~Zhang, C.-C. Chiu, B.~Zoph, E.~D. Cubuk, and Q.~V. Le,
  ``{SpecAugment}: A simple data augmentation method for automatic speech
  recognition,'' in \emph{Proceedings of Annual Conference on the International
  Speech Communication Association (INTERSPEECH)}, 2019.

\bibitem{hori2017joint}
T.~Hori, S.~Watanabe, and J.~Hershey, ``Joint {CTC}/attention decoding for
  end-to-end speech recognition,'' in \emph{Proceedings of Annual Meeting of
  the Association for Computational Linguistics (ACL)}, 2017.

\bibitem{sennrich2016neural}
R.~Sennrich, B.~Haddow, and A.~Birch, ``Neural machine translation of rare
  words with subword units,'' in \emph{Proceedings of Annual Meeting of the
  Association for Computational Linguistics (ACL)}, 2016.

\end{thebibliography}
